\documentclass[11pt]{article}
\usepackage{graphicx}
\usepackage{epsfig}
\usepackage{epsf}
\usepackage{amssymb,amsthm,amscd, amsbsy, array}
 \usepackage{amsmath,bm}
\def\vq{{\bf q}}
\def\vr{{\br}}
\def\lf{\left(}
\def\rg{\right)}
\def\lq{\left[}
\def\rq{\right]}
\def\lgr{\left\{}
\def\rgr{\right\}}
\newcommand{\half}{{\scriptstyle{\frac{1}{2}}}}

\def\cA{{\cal A}}

\def\cR{{\cal R}}

\def\cL{{\cal L}}

\def\vB{{\bB}}
\def\vE{{\vec{E}}}

\def\p{{\partial}}
\def\vx{{\vec x}}

\def\vp{{\vec p}}

\def\beq{\begin{equation}}
\def\eeq{\end{equation}}
\def\beqa{\begin{eqnarray}}
\def\eeqa{\end{eqnarray}}
\def\nn{\nonumber}

\def\smallover#1/#2{\hbox{$\textstyle\frac{#1}{#2}$}} %

\let\ssection=\section
\renewcommand{\section}{\setcounter{equation}{0}\ssection}
\def\vq{{\vec q}}
\def\vp{{\vec p}}

\def\vq{{\vec q}}
\def\vr{{\vec r}}

\def\vv{{\vec v}}

\def\vx{{\vec x}}

\def\vB{{\vec B}}

\def\vg{{\vec g}}

\begin{document}

\title{Dynamics in the magnetic/dual magnetic monopole}

\author{Luigi Martina
\\Dipartimento di Fisica - Universit\`a del Salento\\
Sezione INFN di Lecce\\ Via Arnesano, CP.193,
I-73100 LECCE (Italy)}

\maketitle

\begin{abstract}
Inspired by the geometrical methods allowing the introduction of mechanical
systems confined in the plane and endowed with exotic galilean symmetry,
we resort to the
 Lagrange-Souriau 2-form formalism, in order to look for a wide class
of 3D systems,   involving   not commuting
and/or not canonical variables, but possessing geometric as well gauge symmetries in position and  momenta space too. As a
paradigmatic example,  a charged particle simultaneously interacting with a magnetic
 monopole and a dual  monopole in momenta space  is
considered. The main features of the motions, conservation laws and the analogies with the planar case are discussed.
Possible physical realizations of the model are proposed.
\end{abstract}
\vspace{5mm}
\noindent
\section{Introduction}


 Originated by Lagrange  and continued by
Cartan \cite{Cartan}, the  geometrical formulation  of the
calculus of variations  consists  in   mapping the Lagrangian
function $L(\vx,\vp,t): TQ\times \mathbb{R} \rightarrow
\mathbb{R}$  into the so-called Cartan 1-form $\lambda$ over the
 evolution space  $TQ\times \mathbb{R}$  and minimizing the
corresponding action integral  \beqa \int_\gamma
L(\gamma(t),\dot{\gamma}\lf t \rg,t)dt =
\int_{\widetilde{\gamma}}\lambda  \quad \textrm{with} \quad
\lambda=\frac{\p L}{\p p_i}dx^i+ \left(L-\frac{\p L}{\p
p_i}p_i\right)dt, \label{standardVp}\eeqa where
$\widetilde{\gamma}=(\gamma(t),\dot{\gamma}\lf t \rg,t)$ is the
lifted world-line in the evolution space
  $TQ\times \mathbb{R}$ \cite{SSD}.
The exterior derivative  of the Cartan form provides us with
the closed \textit{Lagrange-Souriau} 2-form $\sigma = d\lambda$.
 The associated Euler-Lagrange equations can be
expressed by looking for the kernel of the 2-form
 $ \sigma(\delta{\widetilde{\gamma}}, \delta {\vec y})=0  $, for any arbitrary movement
  $\delta {\vec y}$ in the evolution space. If the
kernel is one-dimensional, i.e. the rank of $\sigma$ is $2 d =
{\rm{dim}}\lf  TQ \rg$, the variational problem is called regular,
otherwise  it is said  singular and it is  required to resort to
the symplectic reduction techniques in order to describe the evolution space
 foliation in terms of ODE's only
\cite{Cartan,SSD,FaddJ,Mardsen,Marmo}.

Conversely, again following Souriau \cite{SSD}, a generalized
mechanical system is defined  by postulating the existence of a
closed 2-form $\sigma$ on the  evolution space , possessing
constant rank $2 d$. Then, its kernel defines an integrable
foliation with $2 d$-dimensional leaves,  which can be viewed as
generalized solutions of the variational problem. Moreover, by the
Poincar\'e lemma,  $d\sigma=0$ implies   the existence of a Cartan
$1$-form $\lambda$ only locally. In a local chart $\lf {\cal U}
\subseteq TQ\times \mathbb{R}, \xi^1, \ldots, \xi^{2d+1}   \rg$ one
can rewrite $\lambda$ as as $\lambda=a_\alpha d\xi^\alpha$ and
plainly define  a local first-order Lagrangian function  as \beq
\cL= a_\alpha\dot{\xi}^{\alpha} \quad \; \textrm{such that}\quad
\; \int_{\widetilde{\gamma} \subset {\cal U} }\,\lambda =
\int_{\gamma \subset {\pi\,\cal U} } \cL dt
.\label{1-order-action}\eeq  Thus, for those
 models we do not have a usual Lagrange function like in (\ref{standardVp})
defined on the tangent bundle. Put in another way, the
positions coordinates do not satisfy a second-order Newton equation. Moreover,  Lagrangians of the type $\cL$ in different intersecting charts ${\cal U} $ , $ {\cal U}'$ are related by patching procedures involving suitable  transformations.
   The
general question  of the existence of Lagrangian has  been
discussed in \cite{HPAJMP}. Moreover,  if  the Lagrange-Souriau
2-form $\sigma$
   can be split   into a
\textit{symplectic} and a \textit{Hamiltonian part} \cite{SSD}
$\sigma=\omega - dH\wedge dt, \label{sigma-split}$ where
$\omega$ is a closed and regular 2-form on the \textit{phase
space} $TQ$ and $H$  is a Hamiltonian function on $TQ \times
\mathbb{R}$,
 than the equations of motion read $
\omega\big(\dot{\widetilde{\gamma}}\big) =d\,H \label{sELeq}$.
Because of the regularity of  $\omega$,  one introduces the co-symplectic matrix
$(\omega^{\alpha\beta})$  ( i.e.
$\omega^{\alpha\beta}\omega_{\beta\gamma}=
\delta^{\alpha}_{\gamma}$), in such a way that the
Poisson brackets  are defined  by $
\{f,g\}=\omega^{\alpha\beta}\p_\alpha f\p_\beta g
\label{Poisson}$ and  the (Hamilton) equations of motion become
$\dot{\xi}_i=\{\xi_i, H\} $. In the case of singularity of $\omega$, again
symplectic reductions have to be worked out.

On the other hand, in Souriau's framework, one can state the  so
called  inverse problem of the calculus of variations for a given
set of  equations of motion, describing a one-particle system in
presence of a position-dependent force field $\vE$ only. The set
of equations can be  rewritten as the set of 1-forms on $ TQ
\times \mathbb{R}$
\begin{eqnarray}
\alpha_1 = d \vr-\vv\; d t ,\qquad \alpha_2 =  \; m d  \vv - \vE \;dt
,\label{set1form}
\end{eqnarray}
the kernels of both of them provide  the intersections of a set of
hyperplanes in the  evolution space.   Such an
intersection  is described also by the kernel
of the  2-form obtained by the exterior product
  $ \alpha_1 \wedge \alpha_2 \lf\delta y, \delta y'\rg =
\sigma\lf\delta y,  \delta y' \rg = 0 \;$.  In presence of an
electromagnetic field $\vE \lf \vr, t\rg,\vB\lf \vr, t\rg$ acting
on a charge $e$,  Souriau \cite{SSD} generalized the previous
simplest 2-form to
\begin{eqnarray}
\sigma &=& \lf m d \vv- e \vE dt \rg \wedge \lf d \vr- \vv dt \rg
+ e \vB \cdot d\vr \times d \vr, \label{Sem2}
\end{eqnarray}
where we have defined $ \lf d\vr \times d \vr \rg_{k} =
\frac{1}{2} \epsilon_{k i j} dx_i \wedge dx_j $. Then, the usual
equations of motion of a charged particle in the electromagnetic
field are seen to arise as the kernel of $\sigma$, together with
the closure condition $d\sigma=0$,   leading to the  the
homogeneous Maxwell equations for the $\vE$  and $\vB$ fields.
These formulas can be readily generalized to the multi-particles
case.

Now, in the same spirit
we would like to write down a  Lagrangian 2-form for a particle of
mass $m$, which is  subjected both to the electromagnetic field
$\lgr \vE\lf \vr,t \rg , \;\vB\lf \vr,t \rg \rgr $ and  to a
peculiar "environment", in the sense that the relation between
$\vp$ and $\vv$ is more general than how much it was described before.

The simplest example of such a situation is the \textit{exotic}
mechanical model in the 2-dimensional plane proposed by \cite{DH}.
It is defined by a  generalization of the  (\ref{Sem2}) form,
precisely we introduce
\begin{equation}
\sigma = dp_i\wedge dx_i+
\theta\,dp_1\wedge{}dp_2
+eB\,dx_1\wedge dx_2 - d\lf \frac{\vec{p}{\,}^2}{2m} + e V\rg \wedge
dt  , \label{Sourcoup}
\end{equation}
where $\vp =m \, \vv$, $B\lf \vr\rg$ is the magnetic field,
perpendicular to the plane,  $V\lf \vr\rg$  the electric potential
(both of them assumed to be time-independent for simplicity) and
$\theta$ is a constant, called \textit{the non-commutative
parameter}, for reasons clarified below. The resulting equations
of motion read \cite{DH}
\begin{equation}
\displaystyle m^*\dot{x}_{i} = p_{i}-\displaystyle
em\theta\,\epsilon_{ij}E_{j},\qquad \displaystyle \dot{p}_{i} = e
\,E_{i}+e \,B\,\epsilon_{ij}\dot{x}_{j},
\label{DHeqmot}
\end{equation}
where  we have introduced the \textit{effective mass} $
m^*=m\;(1-e\;\theta \;B). \label{effmass} $ The physical novelties are: i) the anomalous velocity
term $-\displaystyle e m \theta\,\epsilon_{ij}E_{j}$, so that
$\dot{\vr} \nparallel \vp$, ii) the derivative of the kinetic
momentum $\vp$ is still determined by the Lorentz force, iii) the
interplay between  $\theta$ and $B$ field in $m^*$. The  2-form (\ref{Sourcoup}) can obtained by
exterior derivation    of the  Cartan 1-form
\begin{equation}
\lambda =  ({p_i}-{ A_i}\,) d{ x_i} -\lf \frac{\vec{p}{\,}^2}{2m}
+ e \, V \rg \,dt + \frac{\theta}{2}\,\epsilon_{ij}\,{ p_i}\, d{
p_j} \label{lambaDH}
\end{equation}
 defining the
action functional as in (\ref{1-order-action}) and involving the
vector potential components $A_i$. Notice that $\lambda$ is gauge
dependent, in contrast with the 2-form $\sigma$, in which only
observable (i.e. gauge invariant) quantities appear.  Furthermore, for $m^*\neq 0$
(\ref{Sourcoup}) can be split in hamiltonian form, leading to the  Poisson
brackets  \begin{equation}
\begin{array}{lll}
\{x_{1},x_{2}\}= \displaystyle\frac{m}{m^*}\,\theta,   \quad
    \{x_{i},p_{j}\}=\displaystyle\frac{m}{m^*}\,\delta_{ij}, \quad
    \{p_{1},p_{2}\}=\displaystyle\frac{m}{m^*}\,eB ,
\end{array}
\label{exocommrel}
\end{equation}
which  satisfy the Jacobi identity for any $B$ field. When  the
effective mass vanishes, i.e. when the magnetic field takes the
critical value $ B_{crit}=\frac{1}{e\theta} $,  the system becomes
singular. Then, the symplectic reduction procedure  leads to a
two-dimensional system characterized by the remarkable Poisson
structure $ \{x_1,x_2\} = \theta $, reminiscent of the
``Chern-Simons mechanics'' \cite{DJT}. Thus, the symplectic plane
plays, simultaneously, the role of both configuration and phase
space. The only motions are those following the Hall law
$p_i = \varepsilon_{i j} \frac{E_j}{B_{crit}}$. Moreover, in the
quantization of the reduced system, not only the position
operators no longer commute, but the quantized equation of motions
yields the Laughlin wave functions \cite{QHE}, which are
the ground states in the Fractional Quantum Hall Effect (FQHE).
Thus, one can claim that the classical counterpart of the
\textit{anyons} are in fact the\textit{ exotic} particles in the
system (\ref{DHeqmot}). In the review article \cite{HMS010}
several examples of 2-dimensional models, which generalize  the
form (\ref{Sourcoup}) and the equations (\ref{DHeqmot}) have been
discussed. Here let us recall that the Poisson structure
(\ref{exocommrel})  can be obtained by applying the Lie-algebraic
Kirillov-Kostant-Souriau method for constructing dynamical
systems, possessing  the (2+1)- Galilei group endowed with a
2-fold central extension \cite{Grigore,otherNC}. The two
 cohomological parameters are
the usual mass $m$ and an
 ``exotic''  parameter $\theta$, describing
the non-commutativity of Galilean boost generators
\begin{equation}
\lgr K_1,K_2 \rgr= - \theta m^2 . \label{exorel}
\end{equation}
 In the context of the condensed matter physics, $ \theta$  can be identified with
 a constant Berry curvature, generated by the lattice structure,
  acting on electron wave-packets  \cite{Xiao:2009rm}.
On the other hand, it can be also viewed as a ``non-relativistic
shadow of the spin"  for  a relativistic  particle, by performing
the so-called ``Jackiw-Nair''  contraction \cite{JaNa}. The main result is that the last
term in (\ref{lambaDH}) can be replaced with a suitable
$\vp$-dependent $A_j\lf \vr, \vp \rg \, dp_j$ 1-form, which
provides a convenient curvature in the momentum space, like the
parameter $\theta$ does in the above example.
 \section{The  general model in (3+1)-dim}
Now, let us look for a further generalizations \cite{BlochHam} -
\cite{Blio1} of the 2-form (\ref{Sem2}) with \textit{momentum}
dependent fields.
  Straightforward algebraic considerations  lead to   define
 the manifestly anti-symmetric covariant  2-tensor on
  the   evolution space
\beqa \sigma &=& \lq \lf 1 - \mu_{i}\rg d p_i - e\; E_i\; dt \rq
\wedge \lf dr_i - g_i\; dt \rg + \half \;e \; B_k \; \epsilon_{k i
j} \;dr_i \wedge dr_j  + \nn
\\ & & \half\;
\kappa_k\; \epsilon_{k i  j}\; d p_i \wedge dp_j +   q_k\;
\epsilon_{k  i  j}\; dr_i \wedge dp_j  , \label{Lag2Dsymm} \eeqa
 where we have put into evidence
    the usual Lorentz force contributions as in (\ref{Sourcoup}),
     while  the  3-vectors
     ${\vec g}$, $\vec \kappa$,
     the diagonal $3\times3$-matrix $Q = \textrm{diag}\lf \mu_{i} \rg $
     and  the anti-symmetric  one $\lf Q_A \rg_{i j} = \epsilon_{i j k}\; q_k$
      depend on all
independent variables and have to be determined in such a way that
$d\, \sigma = 0$ ( the ``Maxwell Principle'' by \cite{SSD}) and to have constant rank.
Similarly to the
effective mass concept in solid state physics, by the
 expression  $ 1 - \mu_{i}$  we would like  to distinguish between
the  \textit{bare}  mass, normalized to 1, and a possible "local"
contribution.
 With
respect to the expression (\ref{set1form})-(\ref{Sourcoup}), we
 introduced the 1-form $dr_i - g_i\; dt $, which defines a general
 relation between the
conjugate momentum and the velocity.

The equations of motion can be written as the kernel of $\sigma \lf \delta y, \cdot \rg =0 $ of the
Lagrange-Souriau form, for any
tangent vector $\delta y = \lf \delta \vr, \delta \vp, \delta t
\rg $. Specifically one obtains the equations \beqa
    e \;\delta \vr  \cdot \vE -  \lf 1- Q \rg  \delta \vp \cdot {\vg} &
= &0,\label{eqEn} \\
  \lf 1- Q -Q_A\rg \delta \vp
& = & e \lf  \vE \; \delta t + \;  \delta \vr \times \vB \rg, \label{eqLor} \\
  \lf 1 - Q \rg \;\lf \delta \vr - {\vec g}\; \delta t \rg & = & {
 -\delta \vr \times \vec q} \;-    \delta \vp \times {\vec \kappa}
 \;.
\label{eqParEx} \eeqa  Equation (\ref{eqLor})  can
be solved  for  $\delta \vp$, if the
matrix $M = {\mathbf{1}} - Q - Q_A$ is invertible. Under such an hypothesis together with
${\rm det}\lf 1 - Q \rg  \neq 0$, one
   replaces $\delta \vp$ in the equation (\ref{eqParEx}), finding
   an equation for the position tangent vector $\delta \vr$, i.e.
   \beq  M^* \; \delta \vr = \lf \lf 1- Q\rg {\vec g} + e\; {\cal K} M^{-1} \cdot \vE   \rg \delta
   t, \label{RMotion}
   \eeq where the  effective mass  matrix  is given by \beqa
   M^{\star}=  M + \lf 2 Q_A - e\; { \kappa} M^{-1} {\cal B} \rg  \; ,
   \quad { \kappa}_{i j} = \epsilon_{i j k} \kappa_k\;, \;
    {\cal B}_{i j} = \epsilon_{i j k}
   B_k. \label{EffMass}
    \eeqa
    Both $M^*$ and the eq.  (\ref{RMotion})
   generalize of the expressions obtained in \cite{DH} leading to (\ref{DHeqmot}).
   Singularities in the motion can arise from the vanishing of ${\rm det}\lf M \rg$,
   ${\rm det}\lf 1 - Q \rg $ and  ${\rm det}\lf M^*\rg $.
   However, if it is not the case,
one can solve (\ref{RMotion}) w.r.t $\delta
   \vr$ and show that
\begin{enumerate}
\item equation (\ref{eqEn}) is identically  satisfied independently
from the specific choice for the vector $\vec g$,
\item the equation (\ref{eqLor}) becomes
\beq
 M^*\delta\vp =
\frac{e}{{\rm det}\lf M\rg}\lf  R\; \vE - {\vec g}^{\; T} N {\vB}
\rg \delta t, \label{eqLorMod} \eeq where matrices $ R $ and $N$
have an involved dependency on $m$, $\vB$, $\vec \kappa$, $Q$ and
$Q_A$ to be spelled here. \end{enumerate}
 Since the pfaffian equations (\ref{RMotion})-(\ref{eqLorMod})
can be proved to be integrable, they are equivalent to the simultaneous first order differential equations
 particle position $ \vr $ and for the variable
    $\vp $, which in the hamiltonian formulation (see Sec. \ref{sec_hamilt}) will play
 the role of  particle momentum, assuming $ {\vec g} = \vp$. Moreover, under such a hypothesis,
   the two equations of motion  simplify to
the system (\ref{DHeqmot}) when $Q$ and  $Q_A$ are set to 0.

Now, accepting  as a law of mechanics \cite{SSD} the closure condition  $ d\sigma = 0 $  for  (\ref{Lag2Dsymm}), we  impose the
vanishing of  the coefficients of the independent $3$-forms. It is
quite natural to assume the  limitations: $\p_{p_i} E_j = \p_{p_i} B_j =0 $ . Thus we are  lead
to the equations
 \beqa \p_{r_j} B_j  = 0,  &  \varepsilon_{k i j } \p_{r_i} E_j = -
\p_t  B_k \label{MaxwellHom}
 , \\ \p_{p_j}  \kappa_j  = 0 ,   & \varepsilon_{k i
j} \p_{p_i} \lq \lf 1- \mu_j \rg g_j \rq =  \p_t \kappa_k,\label{eqtk}\\
 \p_t \mu_{i} =  \p_{r_i} \lq \lf 1- \mu_i \rg g_i \rq  ,   &
 \half \varepsilon_{k i j}\;\p_{r_i} \, \lq \lf 1- \mu_j \rg g_j \rq =  \p_t q_k  ,\label{eqtm}
\\ \p_{r_i} \;\mu_{j} =  \varepsilon_{i j k} \p_{r_j} q_k , &
\p_{r_i} \kappa_j = \varepsilon_{i j k}\p_{p_k} \; \mu_{i} +
\p_{p_i}\; q_j - \delta_{i j} \p_{p_k}\; q_k , \label{skewdiagr2}&
\\
 \p_{r_j} \lq \lf 1- \mu_i \rg g_i \rq +  & \p_{r_i} \lq \lf 1- \mu_j \rg g_j  \rq = 0 , \qquad
i\neq j = 1, 2, 3.\qquad &\label{cond_momento}  \eeqa
                          One can observe that the
homogeneous Maxwell equations (\ref{MaxwellHom}) are the only
restrictions on the electromagnetic fields $\lf \vE, \vB \rg$.
Equations (\ref{eqtk}) are the analogs of the previous relations
in the momentum space for the vector-field $\vec \kappa$. For such
a reason, sometimes $\vec \kappa$ is called dual magnetic field.
As we will see in the Hamiltonian formalism, its existence implies
the non-commutativity of  the spatial coordinates.

 If $\vec \kappa$ is non
trivial in time, then a change in the velocity dependence is
induced  for the mass flow $  \lf \mathbf{1}- Q \rg \vg $, as
prescribed by the second equation in (\ref{eqtk}). In its turn,
equations (\ref{eqtm}) say how the particle mass may change in
time. This seems to be a quite unusual situation, but we cannot
discard it at the moment.  On the other hand,  the first set of
three equations in (\ref{eqtm}) has the form of independent
continuity equations, leading to the global conservation law for
the total mass, i.e. $\p_t \lf \sum_i \mu_i\rg + \p_{r_i} \lq \lf
1- \mu_i \rg g_i \rq = 0 $, which however holds separately in
different directions. Also the skew-symmetric contributions to the
mass matrix $M$ may change on time, but  they generate
modification of the mass flux in space, accordingly to the  second
set   of equations in (\ref{eqtm}).

 The equations in
(\ref{skewdiagr2}) are more difficult  to interpret: they provide
consistency relations for both the space and the momentum
dependency among the mass matrix elements and the dual magnetic
field. Putting such expressions into the equation of motion in the
form (\ref{eqLor})-(\ref{eqParEx}),  in a pure axiomatic way one
re-obtains the equations found in the context of the semiclassical
motion of electronic wave-packets in \cite{Xiao:2009rm}.

 For a  particle with constant mass, i.e. for $\mu_i \equiv 0$ and momentum
${p_i}= g_i $, one easily concludes that $\vec \kappa$ and $\vq$
have to be  constants.    A further analysis of the closure
relations (\ref{eqtm})-(\ref{skewdiagr2}) leads to the expression
$\kappa_i = \sum_{j\neq i} \lf x_j \p_{p_j} q_i -x_i \p_{p_j} q_j
\rg + \chi_i$, where the $q_i$'s and   $\chi_i$ 's depend only on
$\vp$  and moreover the divergenceless condition $\p_{p_j} \chi_j
= 0 $    has to be satisfied.

Limiting ourselves to two spacial dimensions and setting
$\kappa_3= -\theta$, we reobtain the model (\ref{DHeqmot}) above.
More generally  a momentum-dependent non-commutativity field
$\kappa_3=\kappa(\vp)$  was considered in the Snyder space
\cite{Snyder}, with  $\kappa = -\frac{\theta}{1 + \theta p^2}
\varepsilon_{i j} p_i r_j$  (and $\mu_i \equiv 0$ ), or  for the
relativistic   spinning
 particle in the plane, with
$\kappa^{\alpha \beta} = \frac{s}{2}
 \frac{p_\alpha \epsilon^{\alpha\beta\gamma}}
 {(p^2)^{3/2}}$   \cite{anyoneqs,HP1}.  In 3 dimensions  it    is
provided the remarkable example of the monopole field in momentum
space ${\vec \kappa} = \theta \frac{\vp}{|\vp|^3}$, which admits
the spherical symmetry and the canonical relations
$\{r_i,p_j\}=\delta_{ij}$, describing the Poisson structure of the
phase space of
 a 0-mass relativistic particle with non-vanishing helicity (photon) \cite{SSD}, \cite{Carinena}.  Its expression  appears to be consistent
with the experimental data reported for the Anomalous Hall Effect
\cite{AHE} and in Spin Hall Effect \cite{SpinHall} and are theoretically discussed in   \cite{BeMo} and
\cite{HorvMonop}.

  Finally, in the singular submanifold of the
   phase space defined by $
 M^*=0
$,
 we need to look at the proper restrictions on
   vector-fields ${\delta \vp}$ and ${\delta \vr}$, in order to
   avoid motions with infinite velocities. Those restrictions generalize the
    {\it Hall law} discussed in the previous Section.


\section{Hamiltonian Structure \label{sec_hamilt}}

The 2-form $\sigma $ in (\ref{Lag2Dsymm}) can be obtained as the
exterior derivative of the Cartan 1-form, \beq \lambda = \lf \vp -
 \overrightarrow{{\cal A}} \rg \cdot d \vr + {\vec{\cal R}}\, \cdot d \vp -  {\cal T}\, dt.
 \label{cartan1}
\eeq  In this formula the field     $\overrightarrow{{\cal A}} \lf
\vr, t \rg$ is the usual electromagnetic potential, such that $\vB
= \nabla_{\vr} \times \overrightarrow{{\cal A}}$, for which we
have postulated to be momentum-independent. The field
$\overrightarrow{{\cal A}} $  admits the usual gauge
arbitrariness. On the other hand, the electric field is given by
$\vE = - \nabla_{\vr} {\cal T} - \p_t \overrightarrow{{\cal A}} $,
where we assume  that ${\cal T}\lf \vr, \vp, t \rg = {\cal E}\lf
\vp, t \rg+ \varphi\lf \vr,  t \rg$, in order to be stuck to the
previous assumption about the dependency of $\vE$. Furthermore,
the field ${\vec{\cal R}} \lf \vr, \vp, t \rg $ defines the dual
magnetic field $\vec{\kappa} = \nabla_{\vp} \times {\vec{\cal R}}
$, the mass flow  $\lf \mathbf{1} - Q \rg {\vec g} = -\nabla_{\vp}
{\cal T} - \p_t {\vec{\cal R}} $,  the mass components $\mu_i =
\p_{r_i} {\cal R}_i $ and $q_k = \p_{r_i}{\cal R}_j = -
 \p_{r_j}{\cal R}_i$ ($k, i, j$  cyclic). Also ${\vec{\cal R}} $
 admits the gauge arbitrariness both in position and momentum variables. Thus the
 the entire evolution space  is decomposed  in patches, on which the
 Cartan 1-form (\ref{cartan1}) defines local  connections, related
 by gauge transformations satisfying the Maxwell-type equations (\ref{MaxwellHom})-(\ref{cond_momento}).
  In particular, expressed in terms of the gauge field ${\cal R}$ , the
  closure relations in (\ref{eqtm}) and (\ref{skewdiagr2})  become
    \beqa &\p_{r_i}\p_{r_j} {\cal R}_k =0 \quad
\textrm{( $i, j, k$ cyclic)},\qquad \p_{r_i}^2  {\cal R}_j = -
\p_{r_i}\p_{r_j}  {\cal R}_i \;\; \lf i \neq j \rg
,\label{constraints} \\ &\p_t \lf \p_{p_i}{\cal R}_j + 2
\p_{p_j}{\cal R}_i  \rg = 0  \qquad i \neq j, \eeqa with no
summation over repeated indices in the first two equations. Due to
the special form we assumed on the force and magnetic fields, the
above restrictions on $\vec{\cal R}$ limit its space-time
dependency, leaving however the gauge freedom with respect the
momentum variables. As it has been elsewhere remarked \cite{HMS010},
it is possible to perform a change of variables leading to
commutative position variables by a point transformation of the
form $r_i \to r'_i = r_i - {\cal R}_i ({\vr}, {\vp})$. However,
the vector field $\vec {\cal R}$ is defined up to a gauge
transformation generated by an arbitrary function on $\lf \vr, \vp
\rg$. Thus, the  meaning of the notion of position  is unclear in
such a context.

If $\p_t \vec{\cA} = \p_t \vec{\cR} \equiv 0 $, it is possible to
split  (\ref{Lag2Dsymm}) in the hamiltonian
 form  $\sigma=\omega-d {\cal T} \wedge dt $,   by introducing on $ TQ =
\lgr \xi = \lf \vr, \vp \rg \rgr $ the symplectic 2-form \beq
\omega = \lf \delta_{i,j}+ \chi_{ij}\,\rg d p_{\;i} \wedge d
r_{\;j} + \frac{1}{2} \lq b_{i j} \, d r_{\;i} \wedge d r_{\;j} +
\kappa_{i j} \, d p_{\;i} \wedge d p_{\;j} \rq \label{TQHam} ,\eeq
where
 \beqa \chi_{i j} =\left\{%
\begin{array}{ll}
     - \p_{r_j} \cR_{i}, & i \leq j \\
  \p_{r_i} \cR_{j} , & i > j \\
\end{array}%
\right.    \quad  b_{i j} = - \varepsilon_{i j k } B_k, \quad
\kappa_{i j} = \varepsilon_{i j k } \kappa_k .
 \label{altriTens} \eeqa
and the  equations of motion  take the Hamilton form. Moreover, the closure of
$\omega $ is assured by that one of $\sigma$, i.e. by the
equations (\ref{MaxwellHom}) - (\ref{cond_momento}), and their
consequences (\ref{constraints}). Then, accordingly to    the
general expression  (\ref{Poisson}),  the space $TQ$ is endowed
with the  Poisson structure expressed by the co-symplectic matrix
\beqa & \omega^{\alpha, \beta} = \Big(1 - \frac{1}{2} {\rm Tr} \lf
\chi ^2 + b  \lf {\bf 1} + 2 \, \chi \rg \kappa \rg\Big)^{-1}
\\
 &  \lgr \lf {\begin{array}{cc} \kappa + \lq  \chi,
\kappa\rq & 0
 \\
0  &  - b  + \lq \chi,b  \rq \\
\end{array}}
\right) +      \lq 1 - \frac{1}{2}
 {\rm Tr } \lf \chi^2 +b  \, \kappa \rg \rq
 \lf {\begin{array}{cc} 0 &  1
 \\
- 1 & 0 \\
\end{array}}     \nn
\right) + \right. \\ &\left. \lf {\begin{array}{cc} 0 &  \lf \chi^2
+ b \,
 \kappa \rg^T
  \label{cosympl}\\
 -  \lf  \chi^2 +b  \, \kappa \rg  & 0 \\
\end{array}}
\right)\rgr,   \nn
 \eeqa
  non degenerate for
$ \sqrt{\det\lf \omega_{\alpha\beta}\rg} =  1-\frac{1}{2}{\rm Tr}
 \lf \chi^2 +b  \lf {\bf 1} +
2 \, \chi \rg \kappa \rg  \neq 0 . $ Such a factor generalizes the
denominators present in the Poisson brackets (\ref{exocommrel}) or
(\ref{EffMass}). Moreover, it crucially appears in the expression
of the invariant phase-space volume, ensuring the validity of
the Liouville theorem.  Finally, notice that the Poisson structure
is determined only by gauge invariant quantities and  brackets
involving position coordinates $r_i$ are in general
non-commutative.  Several examples systems which can be written in the above formalism are given in \cite{HMS010}, in the following we will discuss the double monopole model.

\section{ The double  monopole}
For a charge subjected only  to a monopole  in the
momentum space, of  strength $\theta$,  and to a uniform electric
field $\vE$, the above procedure leads to equations of motion  readily integrable.
\cite{HorvMonop}. The main feature is that the particle suffers
 a shift $\Delta  = \frac{2
\theta}{ p_0} $ in the direction $\vE \times {\vec p}_0$, being
${\vec p}_0$ the initial  linear kinetic momentum.
This is an important result in controlling spin currents
only by electric fields  \cite{AHE}.

Now, it is natural to consider an electric charge simultaneously subject    to a momentum space (or dual)  monopole and to a  magnetic monopole. From theoretical side, this restores  the "symmetry"  lost  in the
  model described in \cite{Carinena}.
   On the other hand one may figure out a concrete realization
   of such a model, on the base of the experimental evidence
   of isolated monopole excitations in the ice-spin compounds  \cite{Mor}, or
   of the proposal \cite{Qi} in a different context. To include also the effects of a dual monopole, one needs to
find a suitable behavior of  the Berry phase got by an electron
wave-packet moving in the lattice, as described in
\cite{Xiao:2009rm}. For brevity, such a system will be called a
double monopole.    The symplectic structure can be ready derived
from expressions     (\ref{TQHam}) or (\ref{cosympl}) and it can be expressed by the co-symplectic form
\beq \omega^{\alpha,  \beta} = \frac{1}{M^*} \lf \begin{array}{cc}
                                                     \theta |\vr|^3 \, \tilde{p} & -|\vr|^3 |\vp|^3 {\bf 1 }  - e \, \theta \lf  \vp \otimes \vr \rg ^T \\
|\vr|^3 |\vp|^3 {\bf 1}  + e \, \theta \vp \otimes \vr   & e \, |\vp|^3 \, \tilde{r}
                                                   \end{array}\rg , \label{cosymplDM}
\eeq
having introduced $e$ for the  charge-magnetic monopole coupling constant,  $\tilde{p}_{i j} = \varepsilon_{i \,j \,k } p_k$, $\tilde{r}_{i j} = \varepsilon_{i \,j \,k } r_k$
and the effective mass  \beq  M^* =  |\vr|^3 |\vp|^3 - e \theta \;\vr \cdot
\vp. \eeq  From (\ref{cosymplDM}) one gets
$\sqrt{\det\lf \omega^{\alpha, \, \beta}\rg} = \frac{|\vr|^3 |\vp|^3}{M^{*}} $.
Then,  the vanishing of the effective mass $M^*$
will provide an anholonomic  constraint to the dynamics.

 Assuming that no other field is present,  a  charged particle of unitary mass in a double monopole
  has the free Hamiltonian  $H =
\frac{|\vp|^2}{2}$, so that  the equations of motion are
 \beqa
 M^* \dot{r}_i &=&
\lf p_i - e \theta \frac{r_i}{|\vp| |\vr|^3}\rg |\vr|^3 |\vp|^3,
\;  \label{doubleI}
\\
 M^* \dot{p}_i &=& e \varepsilon_{i j k} p_j r_k |\vp|^3 .      \label{doubleII}
\eeqa
On the other hand,  the expression
   \beq \vec{j} = \vr \times \vp - \frac{\theta}{|\vp|} \vp -
\frac{e}{|\vr|}\vr\eeq    is conserved during
  the motion. Moreover,
   its components satisfy the commutation
  relations $\lgr j_i, j_j \rgr = \varepsilon_{i, \, j, \, k} j_k
  $, as well as to commute with the Hamiltonian with respect to the  Poisson structure (\ref{cosymplDM}).  Then $\vec{j}$
   is identified with the total angular momentum, involving both the usual expression for such a quantity for a Dirac monopole and for
 a dual monopole, as in \cite{BeMo}, \cite{ZHN}.    The usual
   arguments about the Liouville integrability for a central system
   may be applied here, in order to make the double monopole system  an
   integrable   model. However, the breaking of the Jacobi identities
$\half \varepsilon_{i j k}\lq p_i,\lq p_j, p_k \rq \rq = e\, \delta \lf \vr \rg $   and
$\half \varepsilon_{i j k}\lq r_i,\lq r_j, r_k \rq \rq = \theta\, \delta \lf \vp \rg $ makes  that conclusion questionable.
On the other hand, the above breaking suggest that a quantum treatment of the problem provides
a quantization both of the magnetic charge $e = N_m/2 $ and of the dual one  $\theta = N_d/2 $ \cite{BeMo}.

A remarkable fact is that the modulus of $\vec j$ takes a non vanishing
   minimum, since it results \beq |{\vec j}|^2 = \lf \vr \times \vp \rg^2 + \lf |\theta| + |e| \rg^2
- 4 |e\, \theta| \,\left\{                                                                                                                     \begin{array}{ll}
                 \textrm{sign}\lf e \theta \rg = 1, & \sin^2\\
\textrm{sign}\lf e \theta \rg = -1, & \cos^2\end{array}
\right.\lf \frac{\widehat{\vr \, \vp}}{2}\rg, \eeq which is achieved in particular when $ \vp \, \|\, \vr $. Certainly,
  it is not surprising that a 0-mechanical angular momentum   {\it electric  charge - magnetic
  monopole}  system possesses a non vanishing total angular momentum
  \cite{Thomson}, but here, a part from the presence of both
   coupling constants, the state is achieved dynamically along particular trajectories,
  intersecting  in the phase space the 5-dimensional unbounded sub-manifold of the vanishing
  effective mass, say $\cal M$. They can exist, since
    the  invariant manifold of constant  total
 energy is
compact only in the momentum subspace, but  the
constant angular momentum sub-manifold may do not  intersect  $\cal
M$, since it depends both on the position and the linear momenta
variables. Notice that the Poisson brackets $\lgr M^*, H \rgr$ are
not vanishing, but restricting them on $\cal M$ one obtains \beq \lgr
M^*, H \rgr|_{\cal M} = 3 |\vp|^2 \lf |\vr|^4 |\vp|^4 - e^2
\theta^2 \rg \vp \cdot \vr. \eeq This relation and $M^*$ itself vanish
consistently with the parallelism condition \beq p_i =
\textrm{sign}\lf e \theta \rg \sqrt{|e \theta |}
\frac{r_i}{|\vr|^2},\label{reduction1}\eeq
 which is dictated by the equations of motion
(\ref{doubleI})-(\ref{doubleII}). Thus a particle trajectory which
reaches the manifold $\cal M$ will remain confined on it.
In this limit equation (\ref{doubleII}) is identically satisfied,
while  equation (\ref{doubleI}) becomes \beq \dot{r}_i = \textrm{sign}\lf e \theta \rg \sqrt{|e \theta |}
\frac{r_i}{|\vr|^2}, \label{redvelocity}\eeq noticing that now the velocity and linear momentum are now proportional (actually equal, having set the mass unitary). Moreover total angular momentum takes the value
$\vec{j} =  - \textrm{sign}\lf e \rg \lf |\theta| +  |e|\rg \frac{\vr}{|\vr|}$, which has minimal modulus. Then, for any initial data $\lgr \vr_0, \, \vp_0\rgr $, such that $| {\vec j}_0| = |\theta| + | e |$,  there exists a position  directed as  $\frac{\vr_{cr}}{|\vr_{cr}|} = - \textrm{sign}\lf e \rg \frac{{\vec j}_0}{|\theta| + | e |}$ along which the constraint (\ref{reduction1}) can be satisfied for the first time.   Since in such a position, the   Hamiltonian takes the value  $H|_{\cal M}  = \frac{|e \theta |}{2 |\vr_{cr}|^2} = E_0$, being $E_0$ its value at the initial point, one concludes that
$ |\vr_{cr}| = \sqrt{\frac{| e \, \theta |}{2 E_0}}$, or \beq \vr_{cr} = - \textrm{sign}\lf e \rg \frac{\sqrt{|e \theta |}}{|\theta| + |e |}\frac{\vec{j}_0}{\sqrt{2 \, E_0}}. \eeq
 Substitution into (\ref{redvelocity}), one gets the velocity $\dot{\vr}_{cr} = \sqrt{2 E_0} \frac{\vr_{cr}}{|\vr_{cr}|}$ of the particle, which from now will follows a restricted dynamics.
\begin{figure}[h]
\includegraphics[width=30pc]{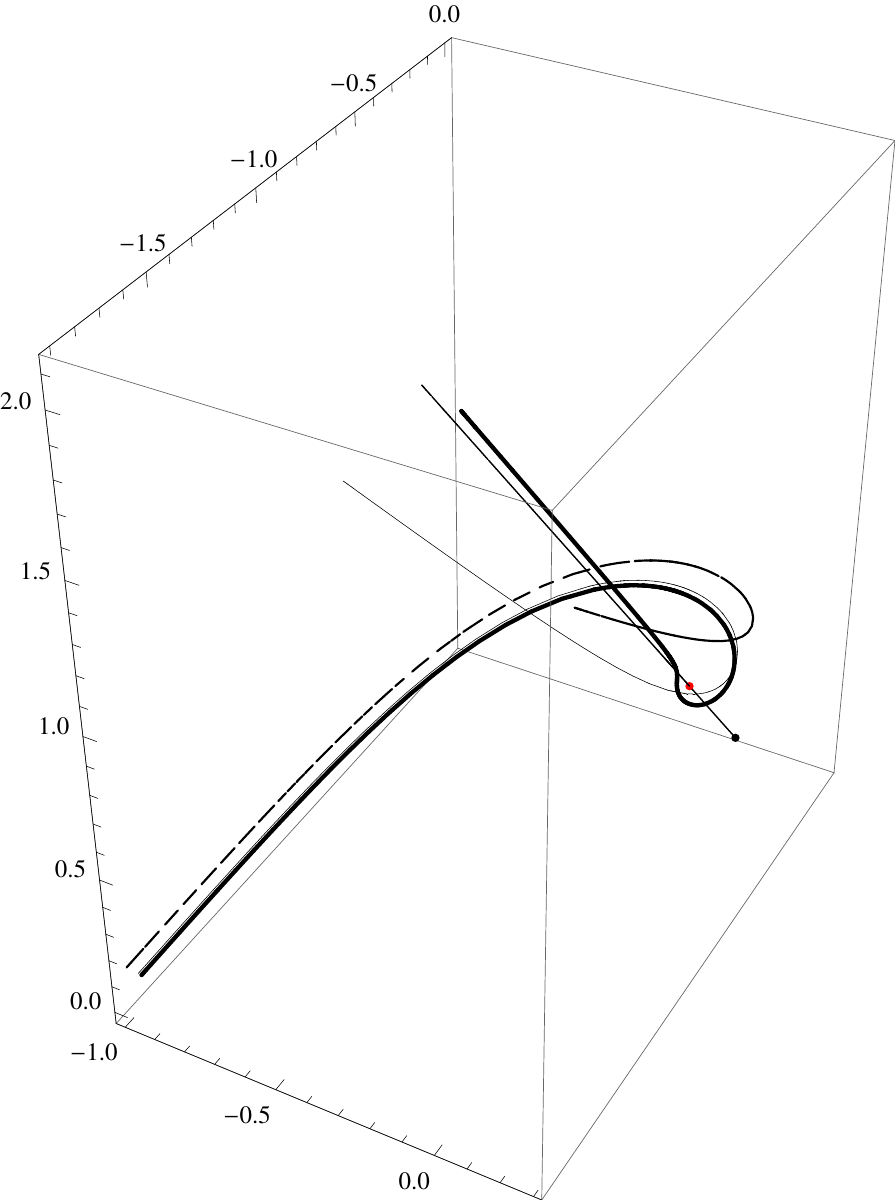}
\begin{minipage}[b]{36pc}\caption{\label{label} Detailed view of the scattering of three charges on a double monopole, showing the details of the trajectory for a particle which is close to be captured in the 0-effective mass sub-manifold ( thick curve), in comparison with two others of increasing angular momentum (thin continuous, then dashed curve). The straight line from the origin indicates  the final trajectory of corresponding 0-effective mass particle and the red point the position $\vr_{cr} $.     }
\end{minipage}
\end{figure}
In fact, on the invariant
critical sub-manifold $\cal M$ the symplectic 2-form $\omega$
 reduces to \beq \omega|_{\cal M} = \half
\varepsilon_{i j k} \frac{e}{|\vr|^3} \lf 1+ |\frac{\theta}{e }|\rg r_k
dr_i \wedge d r_j, \label{ridSympl}\eeq which is proportional by the $\frac{e}{|\vr|^3}$ factor to the symplectic 2-form induced on
 $so\lf 3 \rg^*$  by the  $SO\lf 3 \rg$  co-adjoint action. Resorting to the  usual spherical  foliation,  with  induced Poisson brackets $\lgr \vartheta, \varphi \rgr = - \lq    |e|\lf |\frac{\theta}{e}| + 1\rg |\vr|^2 \sin \varphi \rq^{-1} $ for the angular variables,  one obtains trivial equations of motion $\dot{\vartheta} =  \dot{\varphi} = 0$, since the Hamiltonian is only radius dependent. Then, $ \lf \vartheta, \varphi \rg$ will be fixed by $\frac{\vr_{cr}}{|\vr_{cr}|}$, while  the radius will increase linearly  accordingly to
\beq |\vr| = \sqrt{2 E_0}\, t +|\vr_{cr}|  . \eeq
 Thus,  in the  case of  double monopole  this result realizes  the Hall motions described in Sec. {\bf 1}, with a very similar mechanism based on the interplay of magnetic field and its dual counterpart in the momentum space.

\section{Conclusions}
In conclusion,  a wide set of dynamical systems can be derived
from the Lagrange-Souriau 2-form approach in 3-dimensions.
Generalizations to higher number of degrees of freedom seems
straightforward.  We have shown the conditions to assure their
Hamiltonian formulation. From which an analysis for their
integrability properties can be pursued more plainly, by resorting
to standard methods. However,  one should check if the first order Lagrangian formalism
can be  used equally well, as suggested in \cite{HMS010} Sec. 4. In this perspective, it should be  made a remark  in connection with the procedure,
sometime said of the \textit{ minimal addition},   of coupling the
mechanical system to the the electromagnetic field adopted in
(\ref{Lag2Dsymm}).  It is quite different from the usual
\textit{minimal coupling} procedure, which   yields a very
different Poisson structure. In the context of the 2-dimensional
systems  the two formulations were proved to be equivalent under a
classical Seiberg-Witten transformation of electromagnetic fields \cite{HMS010}, but
no results are yet available in 3D and this hole should be filled in future works.
  In the present article   we have discussed in detail the
double monopole system, which has been proved to be
integrable, since  conserved  energy and total angular momentum has been determined.
In particular, the phenomenon of  capture of the
electric charge into the invariant manifold of vanishing effective mass has been described.
This result cannot be found in the study of the scattering of charges by a magnetic monopole, thus
we expect that  the differential cross section will be strongly influenced by that.
Concerning the symmetry algebra of the double monopole it looks to be  $ so\lf 3 \rg \times \mathbb{R}$,
but the existence of a Runge-Lenz type vector (as it happens for other monopole-like systems)
 is under investigation. This can have a great importance in view of a study
of the quantum version of the proposed model, in analogy to the results obtained in \cite{ZHN}.
Finally, in the light of the articles \cite{Plyu95} the double monopole model presented here could be
useful in generalizing the correspondence among charge - monopole systems and spinning particles or anyons.

\subsection*{Acknowledgments}
 The author expresses his indebtedness to P. Horvathy, M. Plyushchay and  P. Stichel for having been introduced to the problem and
 for continuous and stimulating discussions.  The work has been partially supported
 by the INFN - Sezione of Lecce under the project LE41.



\end{document}